\documentclass[12pt,preprint]{aastex}
\usepackage{natbib}
\usepackage{amsmath}
\usepackage{appendix}
\usepackage{subfigure}
\usepackage{booktabs}

\begin{document}

\title{On the torque exerted by a warped, magnetically threaded accretion disk}
\author{Liu-Chang$^{1}$, Xiang-Dong Li$^{1,2,*}$}

\affil{$^{1}$School of Astronomy and Space Science, Nanjing University, Nanjing 210023, China; 604190724@qq.com}

\affil{$^{2}$Key laboratory of Modern Astronomy and Astrophysics (Nanjing University), Ministry of
Education, Nanjing 210046, China; lixd@nju.edu.cn}

\begin{abstract}
Most astrophysical accretion disks are likely to be warped. In X-ray binaries the spin evolution of an accreting neutron star is critically dependent on the interaction between the neutron star magnetic field and the accretion disk. There have been extensive investigations on the accretion torque exerted by a coplanar disk that is magnetically threaded by the magnetic field lines from the neutron stars, but relevant works on warped/tilted accretion disks are still lacking. In this paper we develop a simplified two-component model, in which the disk is comprised of an inner coplanar part and an outer, tilted part. Based on standard assumption on the formation and evolution of the toroidal magnetic field component, we derive the dimensionless torque and show that a warped/titled disk is more likely to spin up the neutron star compared with a coplanar disk. We also discuss the possible influence of various initial parameters on the torque.
\end{abstract}


\section{INTRODUCTION}

\label{sect:intro}

When a magnetized neutron star (NS) is accreting from a surrounding disk, its spin period evolves due to angular momentum transfer and the interaction between the NS magnetic field and the accretion disk. In their pioneering work, \cite{1979ApJ...232..259G,1979ApJ...234..296G} developed a model for the steady-state configuration of the stellar field lines that thread the accretion disk. This model describes that the angular motion relative to the central star and the inward radial drift of the disk plasma can produce additional toroidal magnetic field component. The interaction of the NS magnetic field and the disk plasma leads to a torque acting on the NS, which can be expressed as follows
\begin{equation}
N=N_{0}+N_{\rm mag},
\end{equation}
where $N_0$ is the material torque due to mass accretion
\begin{equation}
N_0=\dot{M}\Omega_{\rm K}(R_0)R_0^2,
\end{equation}
and $N_{\rm mag}$ is the torque resulting from the magnetic field-accretion disk interaction. Here $\dot{M}$ is the mass accretion rate, $R_0$ is the inner radius of the accretion disk, and $\Omega_{\rm K}$ is the Keplerian angular velocity in the disk.
In the cylindrical coordinate $(R, \phi, z)$ centered on the NS, assuming that the disk is located on the $z = 0$ plane and perpendicular to the NS's spin and magnetic axes, one can derive the magnetic torque $N_{\rm mag}$ resulting from the toroidal field component $B_\phi$ generated by shear motion
between the disk and the vertical field component $B_z$,
\begin{equation}
N_{\rm mag} = -\int_{R_{0}}^{\infty}B_{\phi}B_{z}R^{2}dR.
\end{equation}
\cite{1979ApJ...232..259G,1979ApJ...234..296G} showed that Eq.~(1) can be further written in the following form
\begin{equation}
N=N_0n(\omega),
\end{equation}
where $\omega$ is the fastness parameter defined by
\begin{equation}
\omega\equiv\frac{\Omega_{\rm S}}{\Omega_{\rm K}(R_0)}\equiv \left(\frac{R_0}{R_c}\right)^{3/2}.
\end{equation}
Here $\Omega_{\rm S}$ is the angular velocity of the NS,
\begin{equation}
R_c=\left(\frac{GM}{\Omega_{\rm S}^2}\right)^{1/3}
\end{equation}
is the corotation radius ($G$ and $M$ are the gravitational constant and the NS mass, respectively),
and $n(\omega)$ is a dimensionless function
\begin{equation}
n\approx 1.39\{1-\omega[4.03(1-\omega)^{0.173}-0.878]\}(1-\omega)^{-1}.
\label{GL1}
\end{equation}
There exists a critical fastness parameter $\omega_{\rm c}\simeq 0.35$. When $\omega<\omega_{\rm c}$, $N>0$ and the NS spins up;  when $\omega>\omega_{\rm c}$, $N<0$ and the NS spins down. For a constant accretion rate, the NS spin period will evolve to be the equilibrium spin period with $\omega=\omega_{\rm c}$. Since then there have been many theoretical works trying to modify the form of $n(\omega)$ (e.g., \citealt{1987A&A...183..257W,1991ApJ...370L..39K,1992GApFD..63..179C,1995ApJ...449L.153W,1996A&A...307L...5L,2006A&A...451..581D,2007ApJ...671.1990K}). All of them have assumed that the accretion disk is flat and coplanar with the orbital plane. However, superorbital modulations have been detected in a wide variety of both low- and high-mass X-ray binaries (\citealt{2012MNRAS.420.1575K}, for a review), and warped/tilted disks have been proposed to be a viable mechanism to account for the variability properties. \cite{1973NPhS..246...87K} and \cite{1977ApJ...214..550P} first suggested that the radiation pressure from the NS can cause the disk to be warped and precessing, to explain the 35-d cycle in Her X-1. Radiation-induced warping of accretion disks was further developed by \cite{1996MNRAS.281..357P}, \cite{1999MNRAS.308..207W}, and \cite{2001MNRAS.320..485O}. The general idea is that the outer part of the disk intercepts radiation from the central point source, and if this radiation is absorbed and re-emitted parallel to the local normal to the disk surface, the disk experiences a torque from the radiation pressure, so a small tilt can grow exponentially. In other works, the disk was suggested to be warped due to the magnetic pressure (\citealt{1999ApJ...524.1030L}, \citealt{2004ApJ...604..766P}) and the wind from the accretion disk (\citealt{1994A&A...289..149S}). While most of the theoretical works focus on the dynamical evolution of the disk, there is little attention on the torque exerted by a warped disk. The objective of this paper is to explore how disk warping can influence the accretion torque and the spin evolution of the NS.

\section{Model}
\label{sect:Model}

We construct a simplified toy model for the warped disk, which is composed of two parts, the inner part is coplanar with the orbital plane, and the outer part is tilted at an angle $\theta$ shown in Fig.~1. The inner and outer parts connect at the radius $R_{\rm t}$. We assume that the NS's rotation axis coincides with the magnetic axis, perpendicular to the inner disk. We adopt the cylindrical coordinate system $(x, \phi, z)$ with the NS as the origin and the rotation axis as the $z$ axis. The accretion rate in the warped disk is assumed to be the same as that in the coplanar disk.

\begin{figure}
  \centering
  \includegraphics[scale=0.4]{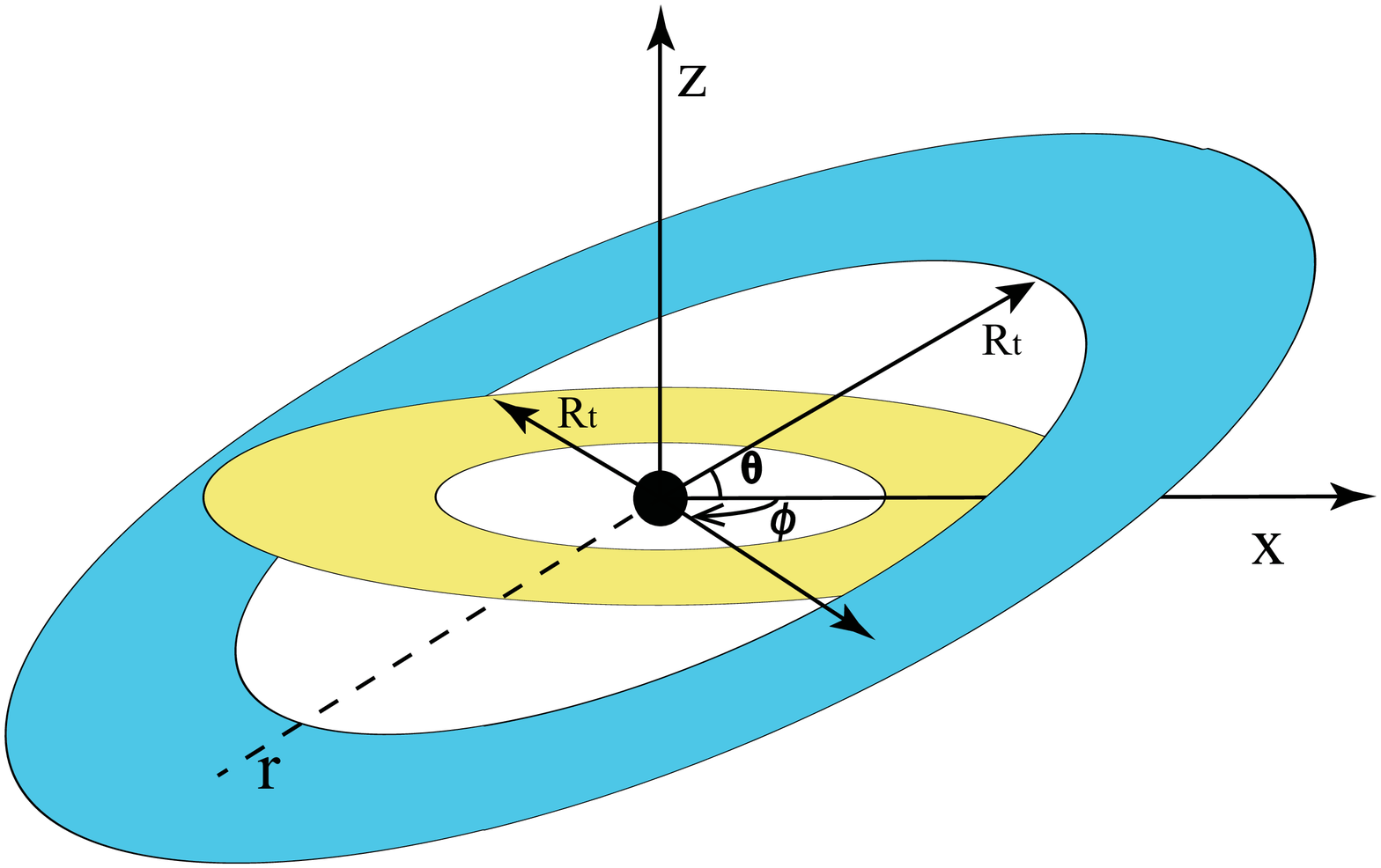}
  \caption{A schematic diagram of the model. The disk is composed of the inner coplanar part and the outer, tilted part, which are indicated with yellow and blue colors, respectively. Note that the two parts should be smoothly connected at $R_t$, which is not explicitly plotted.}
\end{figure}

\cite{1995ApJ...449L.153W} proposed several possible mechanisms for the formation and evolution of the toroidal magnetic field component from the ploidal magnetic field. In this paper, we take the approach of magnetic reconnection outside the disk. The toroidal field component $B_{\phi}$ on the surface of the disk generated by rotational shear is given by
\begin{equation}
\frac { B_ { \phi } } { B _ { z } } = \left\{
\begin{array} { l l }
{ \gamma \left( \Omega_{\rm S} - \Omega _ { \mathrm { K } } \right) / \Omega _ { \mathrm { K } } , } & { \Omega_{\rm S} \leq \Omega _ { \mathrm { K } } } \\ { \gamma  \left( \Omega_{\rm S} - \Omega _ { \mathrm { K } } \right) / \Omega_{\rm S} , } & { \Omega_{\rm S} \geq \Omega _ { \mathrm { K } } },
\end{array} \right.
\end{equation}
where the coefficient $\gamma$ depends on the steepness of the transition between Keplerian motion within the disk and corotation with the star outside the disk (see \citealt{1987A&A...183..257W,1995ApJ...449L.153W}). Its magnitude is around unity to guarantee that $|B_ { \phi }/B _ { z }|$  cannot significantly exceed unity. The vertical component of the magnetic field that crosses the disk is
\begin{equation}
B_{z}=- \frac{\eta \mu }{R^{3}}
\end{equation}
for a dipolar field,  where $\mu$ is the NS magnetic moment and $\eta$ $(\lesssim 1)$ is a screening coefficient.

We first derive the magnetic torque contributed by the inner part of the disk, that is, from $R_0$ to $R_t$. We distinguish two cases when comparing the magnitudes of $R_{t}$ and $R_{c}$. In case 1 with $R_{t} \leq R_{c}$, the torque by the interaction between the NS magnetic field and the inner accretion disk is
\begin{equation}
N_{\rm in,1}=- \int^{R_{t}}_{R_{0}} R ( B_ { \phi }/4 \pi ) \mathbf{B} \cdot\ d \mathbf{S};
\end{equation}
in case 2 with $R_{t} \geq R_{c}$, the torque is
\begin{equation}
N_{\rm in,2}=- \int^{R_ {c} }_{R_ {0} } R ( B _ { \phi }/4 \pi ) \mathbf{B} \cdot\ d \mathbf{S}- \int^{R_ {t} }_{R_ {c} } R ( B _ { \phi }/4 \pi ) \mathbf{B} \cdot\ d \mathbf{S}.
\end{equation}

Combining Eq.~(8)-(11) we have
\begin{equation}
N_{\rm in}=
\begin{cases}
\gamma \eta ^ { 2 } \mu ^ { 2 } ( \frac { 1 } { 3 } R _ { 0 } ^ { - 3 } - \frac { 1 } { 3 } R _ { t } ^ { - 3 } - \frac { 2 } { 3 } R _ { 0 } ^ { -3 / 2 } R _ { c } ^ { - 3 / 2 } + \frac { 2 } { 3 } R _ { t } ^ { - 3 / 2 } R _ { c } ^ { - 3 / 2 } )& {\rm if}\ R_{t} \leq R_{c},\\
\gamma \eta ^ { 2 } \mu ^ { 2 } ( \frac { 1 } { 3 } R _ { 0 } ^ { - 3 } + \frac { 1 } { 3 } R _ { t } ^ { - 3 } + \frac { 2 } { 9 } R _ { c } ^ { - 3 } - \frac { 2 } { 3 } R _ { 0 } ^ { - 3 / 2 } R _ { c } ^ { - 3 / 2 }-\frac { 2 } { 9 } R _ { t } ^ { - 9 / 2 } R _ { c } ^ { 3 / 2 } ) &  {\rm if}\ R_{t} \geq R_{c}.
\end{cases}
\end{equation}
Based on the same approach, we can derive the torque from the outer, warped part of the disk. We assume that the tilted outer disk has an inclination $\theta$ with the horizontal plane, then the magnetic field component vertical to the outer disk is
\begin{equation}
B_{\bot}=-B_{x} \sin \theta + B_{z} \cos \theta =- \cos \theta \frac{\eta \mu }{r^{3}}.
\end{equation}
It's worth noting that $r$ in Eq.~(13) is the radius measured on the inclined disk plane. Because the rotation axis of the NS has an angle $\theta$ with the normal of the outer disk, so the torque parallel to the NS spin is given by
\begin{eqnarray}
N _ {\rm out } & = & -  \cos \theta \int _ { S } r \left( B _ { \phi } / 4 \pi \right) {\bf B} \cdot d{\bf S} \\
            & = &
\begin{cases}
-\cos \theta [\int^{R_ {c} }_{R_ {t} } r ( B _ { \phi }/4 \pi ) \mathbf{B} \cdot\ d \mathbf{S} + \int^{ \infty }_{R_ {c} } r ( B _ { \phi }/4 \pi ) \mathbf{B} \cdot\ d \mathbf{S}] & {\rm if}\ R_{t} \leq R_{c},\\
- \cos \theta \int^{\infty}_{R_{t}} r ( B_ { \phi }/4 \pi ) \mathbf{B} \cdot\ d \mathbf{S} & {\rm if}\ R_{t} \geq R_{c}.
\end{cases} \\
            & = &
\begin{cases}
\gamma \eta ^ { 2 } \mu ^ { 2 } \cos ^ { 3 } \theta ( \frac { 2 } { 9 } R _ { c } ^ { - 3 } + \frac { 1 } { 3 } R _ { t } ^ { - 3 } - \frac { 2 } { 3 } R _ { t } ^ { -3 / 2 } R _ { c } ^ { - 3 / 2 } ) & {\rm if}\ R_{t} \leq R_{c},\\
\gamma \eta ^ { 2 } \mu ^ { 2 } \cos ^ { 3 } \theta ( - \frac { 1 } { 3 } R _ { t } ^ { - 3 } + \frac { 2 } { 9 } R _ { t } ^ { - 9 / 2 } R _ { c } ^ { 3 / 2 } ) & {\rm if}\ R_{t} \geq R_{c}.
\end{cases}
\end{eqnarray}
Note that the limits of the integral is different.

The total torque is then
\begin{equation}
 N= \dot{M}(GM R_{0})^{1/2} + N _ { \rm in }+N _ {\rm out }.
\end{equation}
We assume that $R_{0}=\lambda R_{\rm A}$ where $R_{\rm A}$ is the Alf\'ven radius,
\begin{equation}
R_{\rm A} = \left(\frac{\mu^2}{\dot{M}\sqrt{2GM}}\right)^{2/7},
\end{equation}
the material torque can rewritten to be
\begin{equation}
N_0=\dot{M}(GM R_{0})^{1/2}=\left(\frac{\lambda^{7}}{2}\right)^{1/2}\left(\frac{\mu^{2}}{R_{0}^{3}}\right).
\end{equation}
Then, the dimensionless torque
\begin{equation}
n=
\begin{cases}
1 + \xi\left[ \frac { 1 } { 3 } - \frac { 2 } { 3 } \omega + \frac { 2 } { 9 } \cos ^ { 3 } \theta \cdot \omega ^ { 2 } + \frac { 2 } { 3 } \left( 1 - \cos ^ { 3 } \theta \right) \omega \nu + \frac { 1 } { 3 } \left( \cos ^ { 3 } \theta - 1 \right) \nu ^ { 2 } \right] & {\rm if}\ \omega \leq \nu,\\
1 + \xi \left[ \frac { 1 } { 3 } - \frac { 2 } { 3 } \omega + \frac { 2 } { 9 } \omega ^ { 2 } + \frac { 1 } { 3 } \left( 1 - \cos ^ { 3 } \theta \right) \nu ^ { 2 } + \frac { 2 } { 9 } \left( \cos ^ { 3 } \theta - 1 \right) \frac { \nu ^ { 3 } } { \omega } \right] & {\rm if}\ \omega \geq \nu,
\end{cases}
\end{equation}
where $\xi=2^{1/2}\lambda^{-7/2}\gamma\eta^2$ and $\nu=(R_0/R_t)^{3/2}$.
When $R_{t}\gg R_0$, $\nu\rightarrow 0 $, the whole accretion disk is flat  and coplanar with the orbital plane, Eq.~(20) recovers to
$n=1+\xi(\frac{2}{9}\omega^{2}-\frac{2}{3}\omega+\frac{1}{3})$. When $R_{t}\rightarrow R_{0}$, $\nu\rightarrow 1$, the whole is warped, Eq.~(20) becomes $n=1+\xi(\frac{2}{9}\omega^{2}-\frac{2}{3}\omega+\frac{1}{3})\cos^{3}\theta$.
Note that the magnetic torques in the two cases differ by a factor $\cos^{3}\theta$, which is naturally expected.

\section{Results}
\label{sect:discussion}

Fig.~2 shows the $n-\omega$ relation with $\theta=30\degr$ (left panel) and $60\degr$ (right panel). Other parameters are taken to be $\lambda=0.5$ and $\gamma=\eta^{2}=1$. In each panel, the curves from top to bottom correspond to $\nu=0$, 0.2, 0.4, 0.8, and 1.  When $\nu$ increases, the $n-\omega$ curve becomes flatter. And it is interesting to see that, when $\nu$ is sufficiently large, there is no intersection between $n(\omega)$ and $n=0$, which means that $n$ can be always $>0$ if the warped region is sufficiently large. This tendency becomes more significant for larger $\theta$. This feature is easy to understand, because a tilted outer disk always produces a smaller spin-down torque compared with a coplanar disk.

Fig.~3 shows how the inclination $\theta$ affects the $n-\omega$ relation. In the left and right panels $\nu=0.2$ and 0.8, respectively. Other parameters are same as in Fig.~2. In each panel the curves from top to bottom correspond to $\theta=0\degr$, $30\degr$, $45\degr$, $60\degr$, and $90\degr$, respectively. In the left panel, the change in $\theta$ only slightly influences the $n-\omega$ relation, because the range of the outer tilted disk is relatively small. In the right panel, increase in $\theta$ can obviously flatten the $n-\omega$ relation. In both cases the curves intersect at $\omega=[(3-\sqrt{3})/2]\nu$. At this point $N_{\rm out}=0$, and the total torque is only contributed by the accretion material and the inner disk. It also means that the corotation radius is located in the outer disk, and the spin-up and spin-down torques generated in this region cancel each other. We find that the larger inclination angle is more likely to result in positive torque, \cite{1997ApJ...475L.135W} estimated the torque in the oblique magnetically threaded accretors, and obtained a similar conclusion. The reason is that the spin-down torque comes from the part outside the corotation radius of the disk. When $\theta$ becomes larger, the spin-down torque become smaller, while the spin-up torque is less sensitive to the variation of $\theta$. It is important to see that, with a fixed  $\omega$ (or $\dot{M}$), $n(\omega)$ can be both positive or negative depending on the value of $\theta$. This implies that in a real accreting system, even if the accretion rate is almost constant, the NS may experience spin-up and/or spin-down episodes, which might be related to the torque reversals observed in some X-ray pulsars ( \citealt{1997ApJS..113..367B} ).

Note that we have adopted fiducial values for the parameters $\lambda$, $\gamma$, and $\eta$, all of which are order of unity. Fig.~4 demonstrates their affect on $n(\omega)$.  Here we combine them as a single parameter $\xi= 2^{1/2}\lambda^{-7/2}\gamma\eta^{2}$ and vary its value from 0.1 to 1, and 10. The corresponding $n-\omega$ relation is plotted with the blue, yellow and green curves, respectively. Since this combined parameter determines the magnitude of the magnetic torque, we can see a significant change in the the $n-\omega$ relation, and $n(\omega)>0$ when the contribution from the magnetic torque is relatively small. To see how the magnitude of the torque changes when $\xi$ takes different values, it is more appropriate to evaluate the total torque in units of $\dot{M}(GM R_{c})^{1/2}$ rather $\dot{M}(GM R_{0})^{1/2}$, i.e., $n'=N/\dot{M}(GM R_{c})^{1/2}=n\omega^{1/2}$. Fig.~5 show the $n'-\omega$ relation when $ \xi=0.1$, 1, 3, and 10. The curves in each panel from top to bottom correspond to different values of $\nu$ increasing from 0 to 1. It is obviously seen that disk warping plays a role only when the magnetic torque is sufficiently large. Thus the effect of warping of the outer disk sensitively depends on the structure and evolution of the magnetic field around the inner disk.

\begin{figure}[htbp]
\centering
\subfigure[$ \theta=30^{\circ}$]{
\includegraphics[scale=0.72]{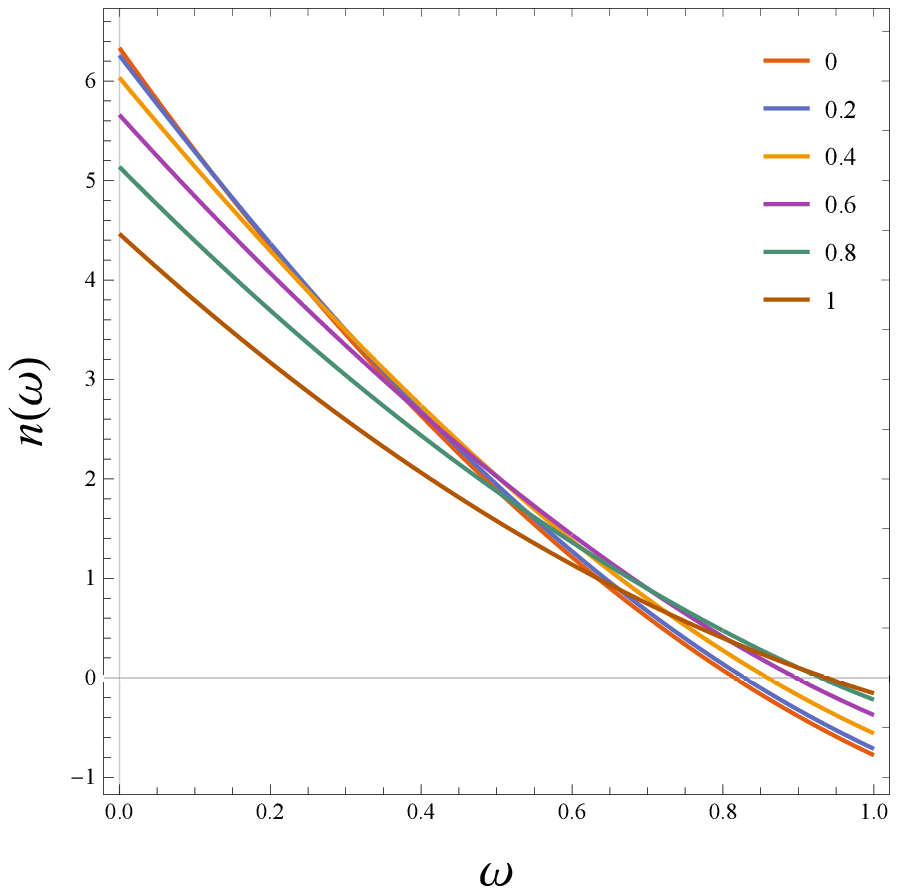}
}
\quad
\subfigure[$ \theta=60^{\circ}$]{
\includegraphics[scale=0.72]{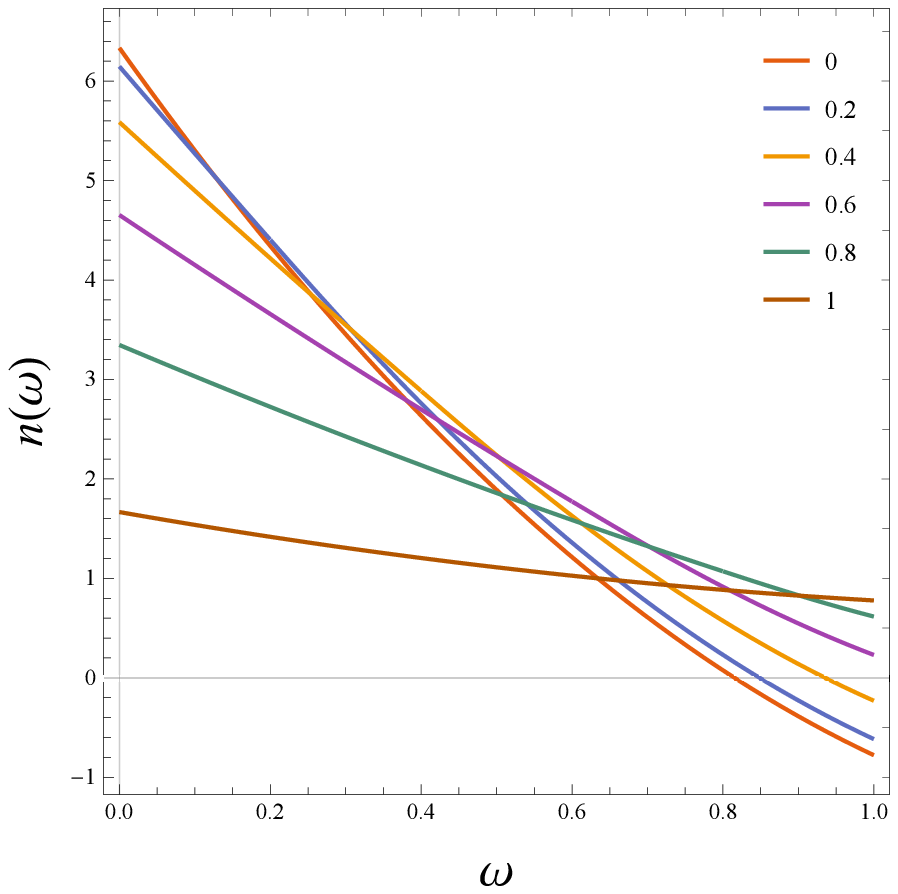}
}
\caption{ The dimensionless torque  $n$ as a function of $\omega$. Shown are the curves with different values of $\nu$. }
\end{figure}

\begin{figure}[htbp]
\centering
\subfigure[$ \nu=0.2$]{
\includegraphics[scale=0.72]{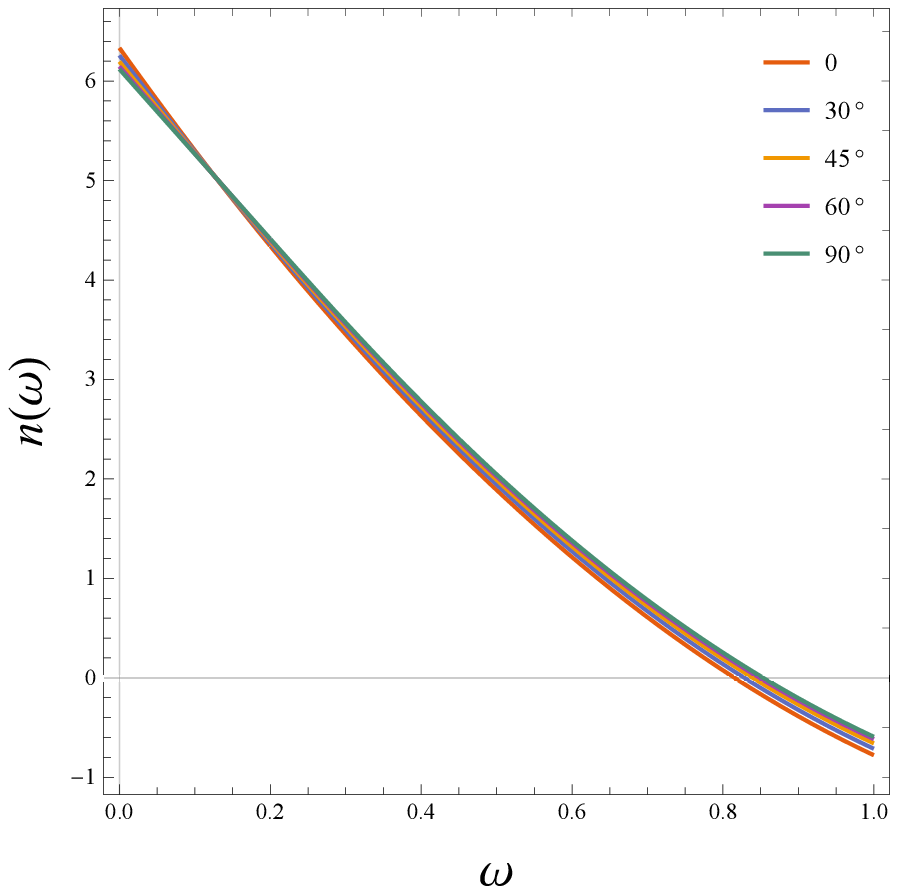}
}
\quad
\subfigure[$ \nu=0.8$]{
\includegraphics[scale=0.72]{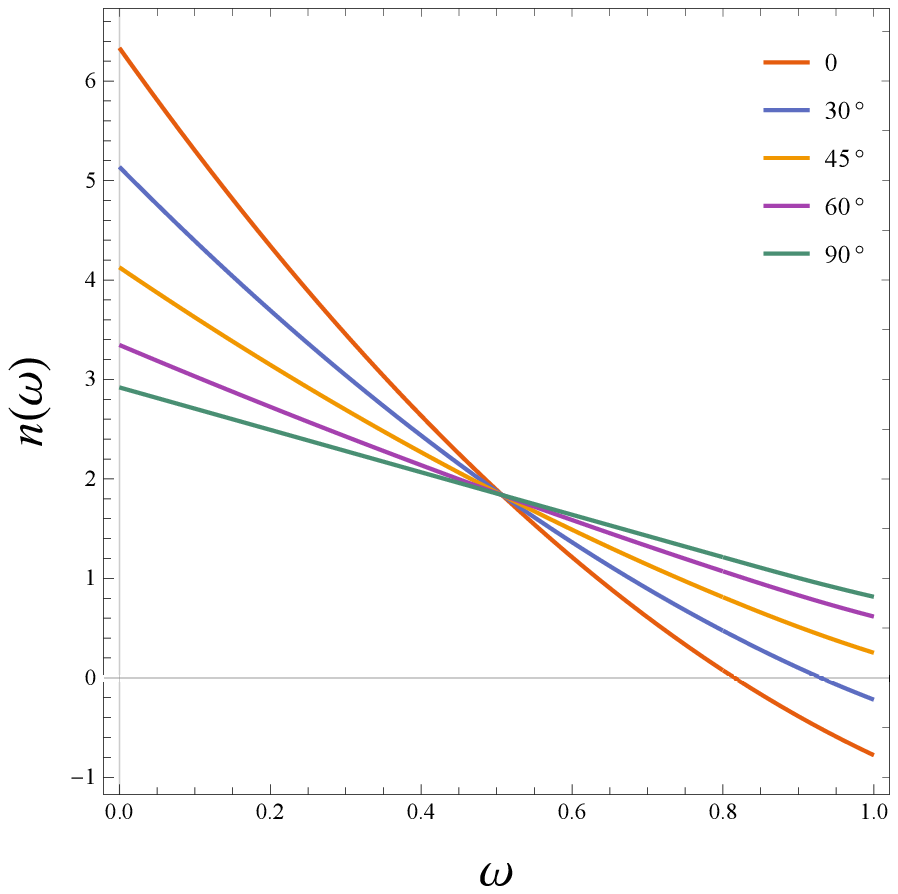}
}
\caption{ The dimensionless torque  $n$ as a function of $\omega$. Shown are the curves with different values of $\theta$. }
\end{figure}

\begin{figure}
   \centering
   \includegraphics[width=8.0cm, angle=0]{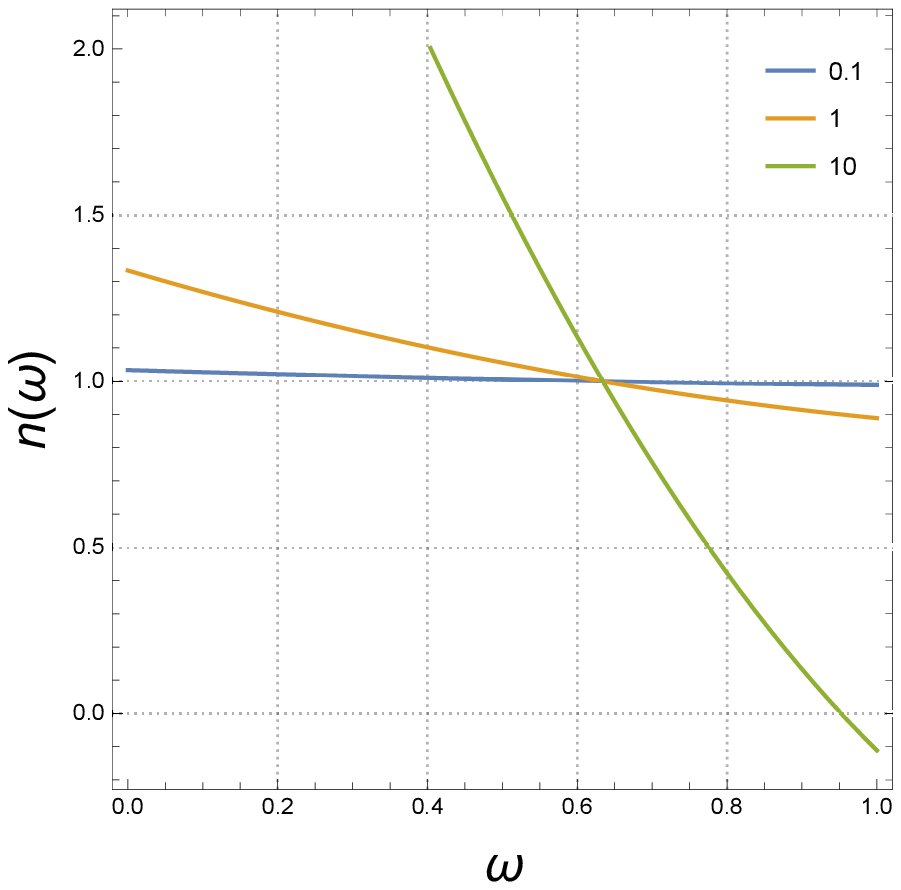}
   \caption{The dimensionless torque $n$ as a function of $\omega$. Shown are the curves with different values of $\xi$. }
   \label{Fig2}
   \end{figure}

\begin{figure}[htbp]
\centering
\subfigure[$\xi=0.1$]{
\includegraphics[scale=0.72]{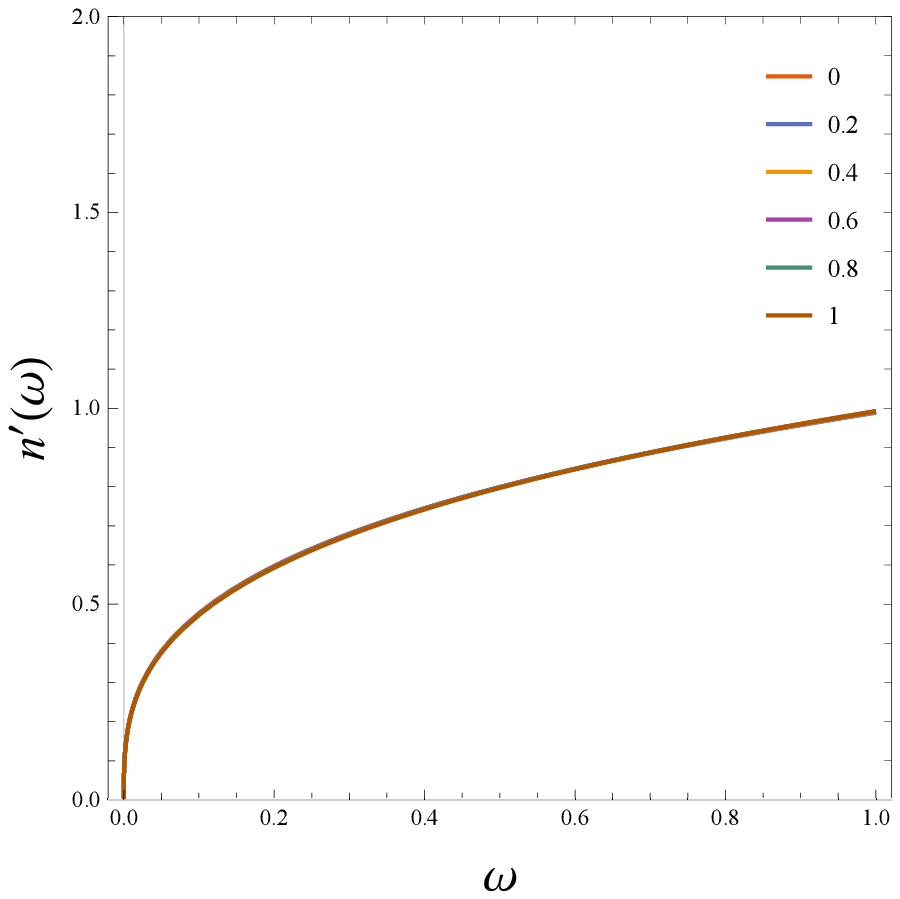}
}
\quad
\subfigure[$\xi=1$]{
\includegraphics[scale=0.72]{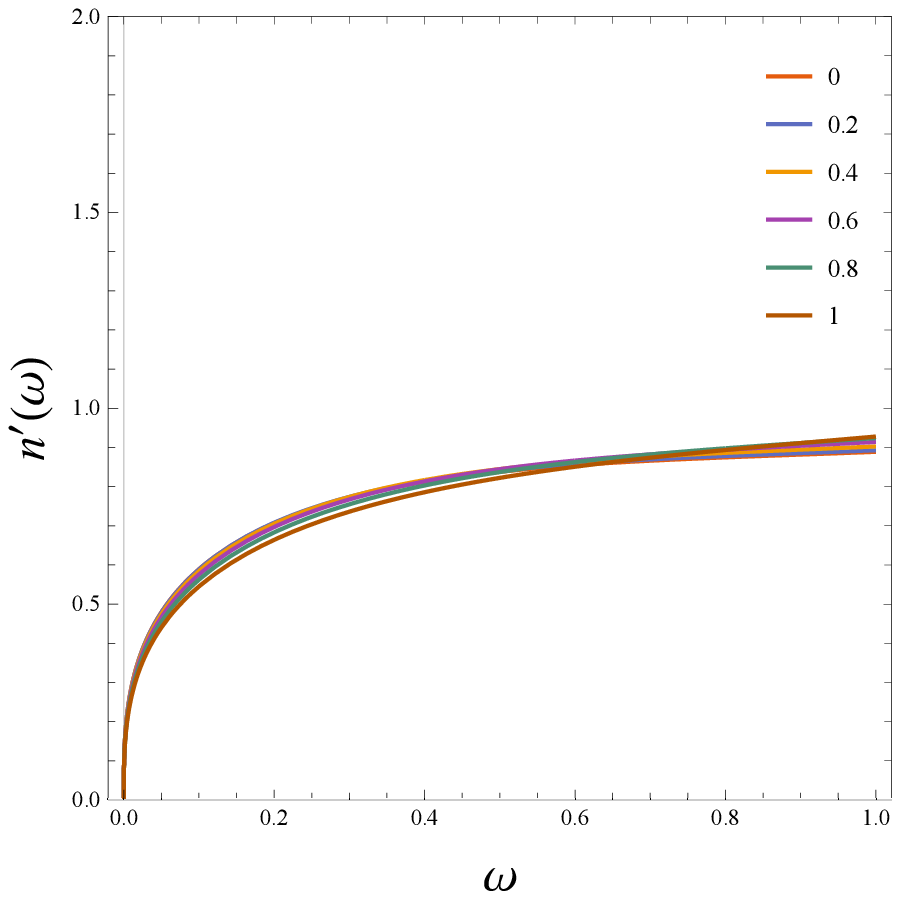}
}
\quad
\subfigure[$\xi=3$]{
\includegraphics[scale=0.72]{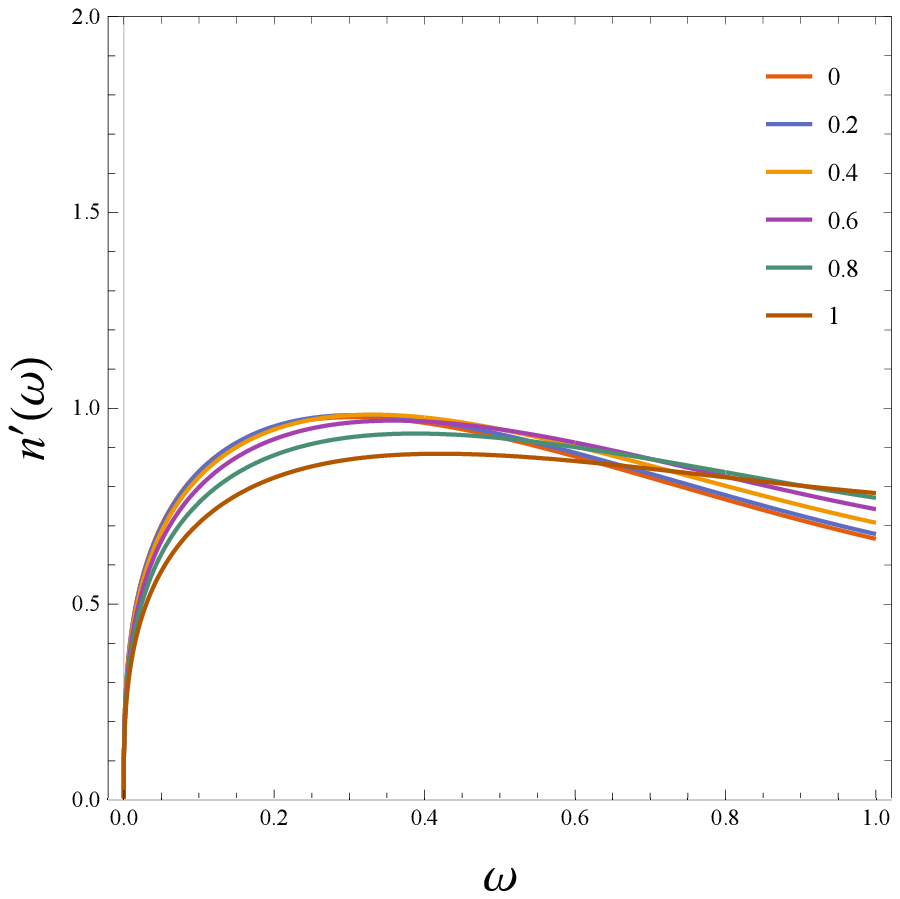}
}
\quad
\subfigure[$\xi=10$]{
\includegraphics[scale=0.72]{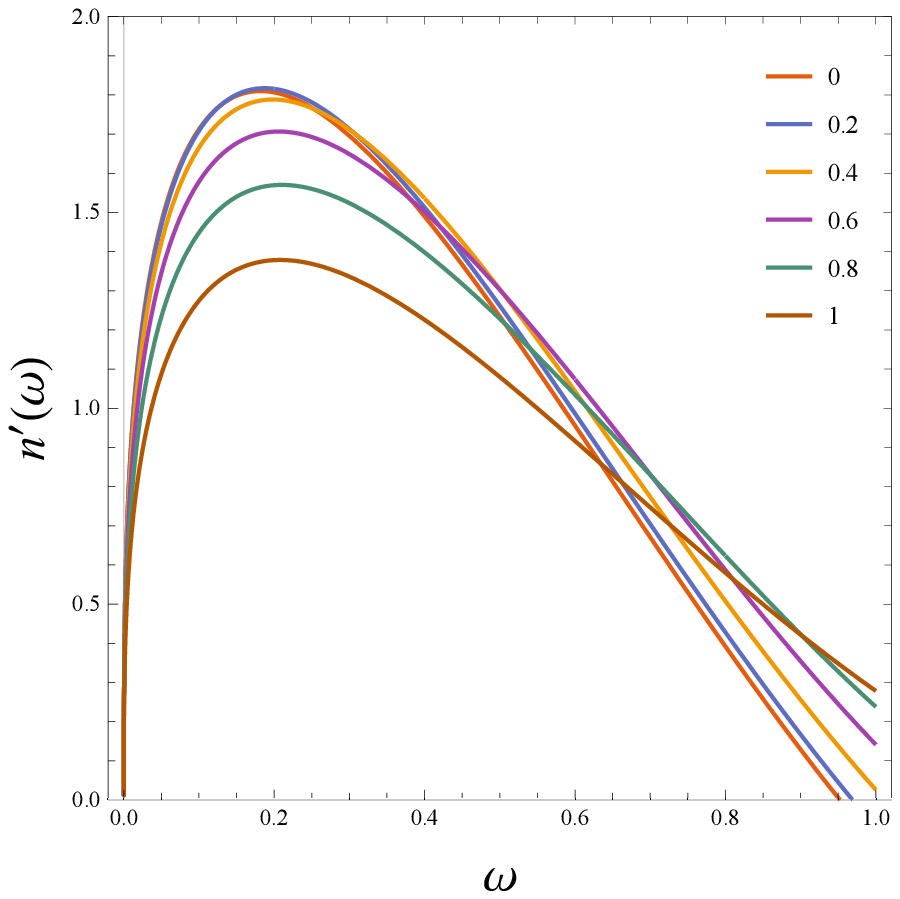}
}
\caption{ The dimensionless torque $n'$ as a function of $\omega$. Shown are the curves with different values of $\xi$ and $\nu$. }
\end{figure}

\section{Discussion and conclusion}

Astrophysical disks are often warped and tilted ( \citealt{2016LNP...905...45N}). In this work we have constructed a simple model to derive the torque exerted by a tilted, magnetized accretion disk on the central NS. The general feature is that a tilted disk is more likely to spin up the NS compared with a coplanar disk, because the spin-down torque decreases
with the inclination. It should be noted that, once part of the disk becomes warped, the warp in the disk propagates both inwards and outwards (\citealt{2019ApJ...875....5M}). In
addition, the radial component of the torque generated by the warped disk
\begin{equation}
N _ {\rm r}= N_{\rm out}\tan\theta=
\begin{cases}
\sin\theta \cos ^ { 2 } \theta\gamma \eta ^ { 2 } \mu ^ { 2 } (\frac { 2 } { 9 } R _ { c } ^ { - 3 } + \frac { 1 } { 3 } R _ { t } ^ { - 3 } - \frac { 2 } { 3 } R _ { t } ^ { -3 / 2 } R _ { c } ^ { - 3 / 2 } ) & {\rm if}\ \omega<\nu\\
\sin\theta \cos ^ { 2 } \theta\gamma \eta ^ { 2 } \mu ^ { 2 } (- \frac { 1 } { 3 } R _ { t } ^ { - 3 } + \frac { 2 } { 9 } R _ { t } ^ { - 9 / 2 } R _ { c } ^ { 3 / 2 }) & {\rm if}\ \omega>\nu.
\end{cases}
\end{equation}
will cause the disk to precess. Thereby, even for a fixed $\omega$, both $\nu$ and $\theta$ will evolve with time. This means that the total torque and hence the
rate of spin change of the NS should vary with time for a constant accretion rate. This feature is not expected in the coplanar disk model.

We finally discuss some caveats in our study. First, we have adopted a simplified disk model which constitute an inner coplanar disk and an outer, tilted disk.
In real situation, the warped disks should be considered as a multi-ring structure, and the change of the inclination in the disk is not stepwise, but continuous.
So the inclination used in our calculation  should be regarded as an averaged one, which reflects the overall condition of the warped and tilted part.
Since $N_{\rm mag}\propto R^{-3}$, the magnetic torque integrated from the warped disk is actually dominated by that from the region close to $R_{t}$, as shown in Fig.~3, this assumption does not significantly influence the final results.

Second, we evaluate the inner radius $R_0$ in terms of the Alf\'ven radius $R_{\rm A}$. An alternative approach is to assume that at $R_0$
the magnetic stresses starts to dominate over viscous stresses in the disk
(e.g., \citealt{1995ApJ...449L.153W}), but this condition suffers the singularity problem when $\omega\rightarrow 1$ (\citealt{2006A&A...451..581D}). Since the torque sensitively depends on the value of $\lambda$ (Eq.~[20]), the uncertainty in $\lambda$ severely limits our understanding of the magnetic field-accretion disk interaction. Early investigations suggest $\lambda\sim 0.5-0.8$ (\citealt{1979ApJ...234..296G, 2005ApJ...634.1214L}), while observations of several X-ray pulsars indicate $\lambda\sim 0.01-1$, varying from source to source (\citealt{2017AstL...43..706F}). \cite{2013MNRAS.433.3048K} performed numerical simulations of accretion onto a magnetized NS and suggested $\lambda\sim 0.77(\mu/\dot{M})^{-0.086}$, which could be more complicated if the magnetic and spin axes of the NS are not aligned (\citealt{2018A&A...617A.126B}). Disk warping may also affect the inner radius of the disk. Because the magnetic stress becomes smaller in the case of warped disk, when considering the conservation of mass and angular momentum in the accretion disk, the inner radius $R_{0}$ will become smaller. See \cite{2018A&A...617A.126B} for similar results for an oblique magnetic rotator.

\acknowledgements
This work was supported by the National Key Research and Development Program of China (2016YFA0400803), the Natural Science Foundation of China under grant No. 11773015 and Project U1838201 supported by NSFC and CAS.

\bibliographystyle{raa}
\bibliography{2020-0238bib}{}

\end{document}